\documentclass[final,5p,twocolumn]{elsarticle}
\usepackage{iftex}
\ifTUTeX
  \usepackage{fontspec}
  \defaultfontfeatures[\rmfamily]{Ligatures=TeXOff}
  \usepackage{amsmath}
  \usepackage[math-style=ISO]{unicode-math}
  \newcommand{\indicator}{\mathbb{1}}
\else
  \usepackage[T1]{fontenc}
  \usepackage{amsmath}
  \usepackage{newtxtext}
  \usepackage[slantedGreek]{newtxmath}
  \usepackage{dsfont}
  \newcommand{\indicator}{\mathds{1}}
  \DeclareUnicodeCharacter{2212}{\textminus}
  \newcommand{\symbfup}[1]{\ifx#1\eta\boldsymbol{\etaup}\else\ifx#1\xi\boldsymbol{\xiup}\else\ifx#1\nu\boldsymbol{\nuup}\else\mathbf{#1}\fi\fi\fi}
  \newcommand{\symbfit}[1]{\boldsymbol{#1}}
\fi
\usepackage{booktabs,float,stfloats}
\usepackage{algpseudocode}
\usepackage[hidelinks,unicode]{hyperref}
\graphicspath{{figures/}}
\biboptions{sort&compress}
\usepackage{etoolbox}
\makeatletter
\patchcmd{\NAT@citexnum}{--\NAT@penalty}{\textendash\NAT@penalty}{}{}
\makeatother

\floatstyle{ruled}
\newfloat{algorithm}{tbp}{loa}
\floatname{algorithm}{Algorithm}

\begin{document}

\begin{frontmatter}

\title{Safety-Aware Bayesian Inference of Personalized Lane Keeping Controllers with Hardware-in-the-Loop Validation}

\author{Beomchang Kim}
\author{Suyong Park}
\author{Kyoungseok Han\corref{cor1}}
\ead{kyoungsh@hanyang.ac.kr}
\cortext[cor1]{Corresponding author.}

\affiliation{organization={Department of Automotive Engineering, Hanyang University},
            addressline={222, Wangsimni-ro, Seongdong-gu},
            city={Seoul},
            postcode={04763},
            country={Korea}}

\begin{abstract}
Conventional lane keeping assistance systems generally do not account for individual differences in lane keeping behavior. We propose a personalized pure pursuit controller whose parameters are inferred from each driver's demonstrations by safety-aware Bayesian inference. The preferred lateral offset varies with road curvature, and the speed-dependent look-ahead distance, which is not directly observed, is modeled using a latent Gaussian process.
The likelihood compares demonstrated trajectories with closed-loop rollouts of the controller, and a soft safety penalty is added to the inference target. Markov chain Monte Carlo samples the controller parameters from the resulting safety-aware posterior, and posterior expectations define the deployed controller.
Synthetic recovery and driving-simulator demonstrations from two drivers showed that the inference recovers a known controller and yields driver-specific preferred paths and look-ahead functions that differ beyond their posterior uncertainty. The safety penalty increased the deployed controller's minimum lane clearance with little change in trajectory fit. The increase was larger in conditions where the posterior without the penalty showed greater uncertainty in clearance.
To verify that the deployed controller works on real hardware, we tested it for both drivers in hardware-in-the-loop simulation with a physical steering wheel. For each driver, it stayed close to its offline rollouts, maintained lane clearance above the warning threshold, and reproduced the driver's held-out trajectories more closely than a baseline controller.
\end{abstract}


\end{frontmatter}

\section{Introduction}\label{sec:introduction}
Advanced driver assistance systems have become common in production vehicles and are intended to improve driving safety and comfort~\cite{bengler}. Lane keeping assistance systems (LKAS) are one of these systems and provide continuous steering control to keep a vehicle on a path within its lane.

However, conventional systems are generally designed around nominal driver behavior, whereas lane keeping behavior varies across drivers~\cite{barendswaard}.
As a result, the assistance may not match a driver's own driving style, and drivers may override it~\cite{park,reagan,romano}.

This mismatch has motivated driver-adaptive assistance using different sources of driver-specific information~\cite{hasenjager}.
For lane keeping, assistance torque has been adapted to driving styles identified through fuzzy classification~\cite{rath}.
Driver-specific steering behavior has also been incorporated into neuro-adaptive lane keeping controllers~\cite{zhou}.
Data-driven driver models have been used to predict unintended lane departures and support personalized assistance~\cite{lefevre}, while driver-specific desired paths and steering-model parameters have been identified from simulator driving data~\cite{schnelle}.
Human-like paths have also been learned from driving data~\cite{lihuman}, while models of lane-offset selection have been fitted to individual drivers and used to classify driving styles~\cite{igneczi2025}.
Other approaches have learned drivers’ assistance preferences more directly.
Bayesian preference learning has been applied to pairwise comparisons of candidate trajectories~\cite{ran}, while naturalistic driving characteristics have been mapped to assistance preferences obtained through driver evaluations~\cite{chen}.
More recently, lane keeping policies have been adapted online through reinforcement learning guided by driver interventions~\cite{yang}.
When personalization relies on driving demonstrations, a further question is how precisely those demonstrations identify the driver’s preferred path and steering response.

Human steering has been modeled through anticipatory and compensatory control~\cite{donges}, and through the visual direction of near and far points on the road~\cite{salvucci}.
Pure pursuit is a geometric path-tracking method that determines steering from a look-ahead point on a reference path~\cite{coulter}.
The reference path specifies the desired lateral position, while the look-ahead distance governs steering responsiveness.
The use of a look-ahead point is also consistent with models identified from human steering data~\cite{vanderel}.
Within this framework, driver-specific look-ahead distances have been learned from driving data for personalized path tracking~\cite{li}.
However, the model takes the reference path as an input and therefore does not infer the driver’s preferred lateral position within the lane.

The controller parameters underlying these behaviors are not directly observable and must be inferred from demonstrations. Driver steering models have been identified from measured steering, for example by fitting a model-predictive steering controller to each driver~\cite{keen}, but the identified parameters are point estimates. Inverse reinforcement learning infers a reward that explains demonstrated driving~\cite{ziebart,kuderer,huang}, and its Bayesian form provides a posterior over the reward~\cite{ramachandran}. In transportation, Bayesian calibration has quantified the parameter uncertainty of car-following models and adaptive cruise controllers~\cite{zhang,jiang}.
Such uncertainty also arises in lane keeping, since different combinations of preferred-path and tracking parameters may reproduce similar trajectories.

Safety in learning from demonstrations is commonly addressed by learning constraints from the demonstrations or by evaluating the learned policy afterwards~\cite{liuconstraint,brown}. In constrained Bayesian inference, hard constraints have been relaxed into exponential penalties on the distance to the constraint set, which favor feasible values while retaining posterior uncertainty~\cite{duan,presman}. To the best of our knowledge, the parameters of a lateral controller have not been inferred as a posterior distribution from demonstrated trajectories with a safety penalty in the inference target.

\begin{figure*}[!tb]
  \centering
  \includegraphics[width=\linewidth]{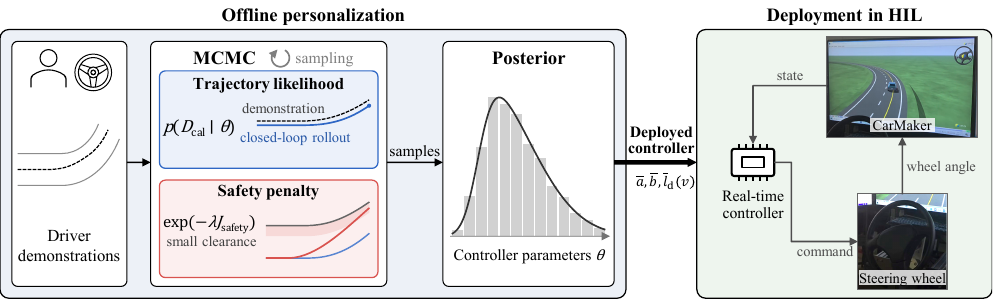}
  \caption{Overview of the proposed framework: offline personalization by MCMC and deployment in HIL.}
  \label{fig:framework}
\end{figure*}

In this paper, we formulate LKAS personalization as safety-aware Bayesian inference of a driver's preferred-path and tracking parameters from demonstrated trajectories.
The controller uses a curvature-dependent lateral offset to represent the preferred path within the lane and a speed-dependent look-ahead distance to govern how responsively that path is tracked. A latent Gaussian process (GP) models the look-ahead distance to capture its speed dependence without fixing a parametric form. Trajectory matching alone, however, does not ensure adequate lane clearance. We therefore incorporate a soft safety penalty on lane clearance, evaluated from closed-loop rollouts of candidate controllers, directly into the inference target.
Markov chain Monte Carlo (MCMC) samples the resulting safety-aware distribution rather than identifying only a single point estimate, revealing the range of plausible controller parameters and their dependencies. Posterior expectations then define the deployed controller, allowing safety costs across plausible parameter values to influence its behavior without prescribing a separate clearance margin for each condition. We evaluate whether the inference recovers a known controller from synthetic data, compare the controllers inferred from two drivers’ demonstrations, and examine how the safety penalty affects Bayesian and deterministic estimates. In hardware-in-the-loop (HIL) tests with a physical steering wheel, we assess how closely the deployed controller reproduces each driver’s held-out trajectories and whether it maintains lane clearance above the warning threshold. Fig.~\ref{fig:framework} summarizes the proposed framework. This paper extends our preliminary work~\cite{kimksae}, in which the preferred offset and a single look-ahead distance were inferred from one driver’s demonstrations at one speed without applying the safety penalty. Here, we also validate the deployed controller in HIL tests.

This paper is organized as follows. Section~\ref{sec:preliminaries} reviews pure pursuit control, the kinematic bicycle model, GPs, and MCMC. Section~\ref{sec:formulation} presents the personalized LKAS formulation. Section~\ref{sec:experiments} reports the synthetic recovery, the calibration results, the comparison between drivers, the effect of the safety penalty, and the comparison with deterministic identification. Section~\ref{sec:held-out} presents the HIL implementation and validation, and Section~\ref{sec:conclusion} concludes.

\section{Preliminaries}\label{sec:preliminaries}

\subsection{Pure pursuit}\label{sec:pursuit}

Pure pursuit controls a vehicle's steering by calculating the curvature needed to reach a designated look-ahead point on the path via a circular arc, which gives
\begin{equation}
  \label{eq:pursuit}
\delta=\arctan \left(\frac{2L\sin \alpha}{l_{\mathrm{d}}}\right)
\end{equation}
where \(L\) is the wheelbase and \(\alpha\) is the angle between the vehicle heading and the line from the rear-axle center to the look-ahead point.
The look-ahead distance \(l_{\mathrm{d}}\) parameterizes the steering response.

\subsection{Kinematic bicycle model}\label{sec:bicycle}

Vehicle motion is described by a standard rear-axle kinematic bicycle model.
With the rear-axle center at \(X,Y\) and heading \(\psi\) in the global frame,
\begin{equation}
\dot{X}=v \cos \psi, \dot{Y}=v \sin \psi, \dot{\psi}=\frac{v}{L} \tan \delta
\end{equation}
Discretizing these dynamics with integration step \(\Delta t\) gives
\begin{equation}
  \label{eq:discrete-bicycle}
\begin{aligned}
X_{t+1}&=X_{t}+\Delta t\,v_{t} \cos \psi_{t} \\
Y_{t+1}&=Y_{t}+\Delta t\,v_{t} \sin \psi_{t} \\
\psi_{t+1}&=\psi_{t}+\Delta t\frac{v_{t}}{L} \tan \delta_{t}
\end{aligned}
\end{equation}

\subsection{Gaussian process}\label{sec:gp}

A GP defines a distribution over functions through a mean function \(m(\cdot)\) and a covariance function \(k(\cdot,\cdot)\)~\cite{rasmussen}. For a finite set of inputs \(u_{1},\ldots,u_{M}\), the corresponding function values \(\symbfup{f}={[f(u_{1}),\ldots,f(u_{M})]}^{T}\) follow a multivariate Gaussian distribution,
\begin{equation}
\symbfup{f}\sim \mathcal{N}(\symbfup{m},K),\quad \symbfup{m}_{i}=m(u_{i}),\quad K_{ij}=k(u_{i},u_{j})
\end{equation}

The covariance function specifies how function values are correlated across inputs and thereby encodes assumptions such as smoothness without fixing a parametric form.
Conditioning on \(\symbfup{f}\) gives a Gaussian distribution at any new input, with mean and variance in  closed form. In this study, a GP is used as a prior for the latent speed-dependent look-ahead function, with its covariance function given in Section~\ref{sec:controller} and its mean function in Section~\ref{sec:data}.

\subsection{Markov chain Monte Carlo}\label{sec:mcmc}

Bayesian inference represents an unknown parameter vector \(\theta\) by its posterior distribution given observed data \(D\).
The posterior follows from Bayes’ rule,
\begin{equation}
p(\theta \mid D)=\frac{p(D\mid \theta)p(\theta)}{\int p(D\mid \theta)p(\theta)d\theta}
\end{equation}
where \(p(D\mid \theta)\) is the likelihood and \(p(\theta)\) the prior.
The integral in the denominator is generally intractable, so the posterior is known up to a constant factor.

MCMC generates samples from a distribution known up to such a constant.
Write \(\pi\) for the target distribution. A Markov chain with transition kernel \(P\) is constructed so that \(\pi\) is its stationary distribution,
\begin{equation}
\pi (\theta ')=\int P(\theta '\mid \theta)\pi (\theta)d\theta
\end{equation}
A sufficient condition is detailed balance,
\begin{equation}
\pi (\theta)P(\theta '\mid \theta)=\pi (\theta ')P(\theta \mid \theta ')
\end{equation}
Under suitable ergodicity conditions, the distribution of the chain converges to \(\pi\).

The Metropolis–Hastings algorithm constructs such a kernel.
Given the current state \(\theta\), a candidate \(\theta '\) is drawn from a proposal density \(g(\theta '\mid \theta)\) and accepted with probability
\begin{equation}
\min \left\{1,\frac{\pi (\theta ')g(\theta \mid \theta ')}{\pi (\theta)g(\theta '\mid \theta)}\right\}
\end{equation}
If the candidate is accepted, the chain moves to \(\theta '\); otherwise it remains at \(\theta\) and the current state is recorded again.
The target appears only as a ratio, so its normalizing constant is not required.

Expectations under \(\pi\) are estimated by averages along the chain,
\begin{equation} 
  \label{eq:expectation}
\frac{1}{N}\sum_{i=1}^{N} h(\theta^{(i)})\to \mathbb{E}_{\pi}[h(\theta)]
\end{equation}
as \(N\to \infty\). An initial warmup period is used for tuning and discarded.
The retained states remain correlated, so their effective information depends on the chain autocorrelation.

\section{Personalized LKAS formulation}\label{sec:formulation}

\subsection{Personalized controller model}\label{sec:controller}

Following the approximately linear relationship between lateral offset and road curvature observed in human driving data~\cite{igneczi}, we model the preferred lateral offset as an affine function of signed road curvature $\kappa$, with curvature and lateral offsets positive to the left:
\begin{equation}
  \label{eq:preferred-offset}
y_{\mathrm{off}}(\kappa)=a+b\kappa
\end{equation}

\begin{figure}[!tb]
  \centering
  \includegraphics[width=\linewidth]{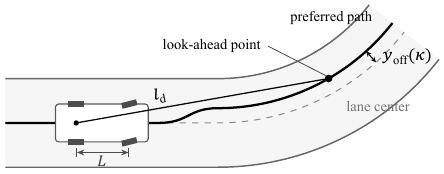}
  \caption{Preferred path and pure pursuit geometry.}
  \label{fig:controller-geometry}
\end{figure}

To avoid a discontinuous shift of the reference path at curve entry, the preferred offset is introduced as \(w(s)(a+b\kappa_{\mathrm{curve}})\), where \(s\) is the distance along the road and \(w(s)\) increases smoothly from zero to one over the transition region.
The reference path therefore follows the lane center before the transition and reaches the offset of Eq.~\eqref{eq:preferred-offset} on the curve (Fig.~\ref{fig:controller-geometry}).

To ensure positivity, we model the speed-dependent look-ahead distance in log space as
\begin{equation}
\log l_{\mathrm{d}}(v)=f(v)
\end{equation}
Let \(D_{\mathrm{cal}}\) denote the driver demonstrations used for inference, and \(v_{1}, \ldots, v_{M}\) denote the \(M\) distinct speeds in \(D_{\mathrm{cal}}\). The corresponding latent function values are denoted by \(\symbfup{f}={[f_{1},\ldots,f_{M}]}^{T}\). We assign the GP prior with squared-exponential covariance:
\begin{equation}
  \label{eq:gp-prior}
\symbfup{f}\mid \varphi \sim \mathcal{N}(\symbfup{m},K_{\varphi}),\quad \varphi=(\sigma_{f},\ell)
\end{equation}
\begin{equation}
  \label{eq:kernel}
{(K_{\varphi})}_{ij}={\sigma_{f}}^{2}\exp \left(-\frac{{(u_{i}-u_{j})}^{2}}{2\ell^{2}}\right)
\end{equation}
where \(u_{i}\) denotes the speed \(v_{i}\) after centering and scaling by 10 km/h. Since a fixed mean function is difficult to specify in advance, the prescribed mean is extended by a level and a slope whose coefficients, with Gaussian priors, are inferred from the data by integrating them out, which adds their prior covariance to \(K_{\varphi}\), and the GP models the deviations from this global trend~\cite{rasmussen}. Hereafter, \(K_{\varphi}\) denotes the resulting covariance. The controller parameters \(\theta=(a,b,\symbfup{f},\varphi)\), with \(\varphi=(\sigma_{f},\ell)\), are inferred from the demonstrations of each driver.

\subsection{Demonstrations and trajectory likelihood}\label{sec:likelihood}

For episode \(n\) in \(D_{\mathrm{cal}}\), the recorded road geometry, longitudinal speed, and initial state define the rollout conditions \(\mathcal{S}^{n}\).
Given \(\mathcal{S}^{n}\) and controller parameters \(\theta\), a closed-loop rollout is generated with the controller and the kinematic bicycle model.
The demonstration and rollout are compared through \(\symbfup{z}_{t}={[e_{y},e_{\psi},\delta]}^{T}\), where \(e_{y}\) and \(e_{\psi}\) are the lateral offset of the rear-axle center and the heading error relative to the lane center, and \(\delta\) is the front-wheel steering angle.
Assuming independent Gaussian observation errors with fixed covariance \(\Sigma\), the trajectory likelihood is
\begin{equation}
  \label{eq:likelihood}
p(D_{\mathrm{cal}}\mid \theta,\mathcal{S})=\prod_{n,t} \mathcal{N}({\symbfup{z}_{t}}^{n};{\widehat{\symbfup{z}}_{t}}^{n}(\theta),\Sigma)
\end{equation}
Thus, the likelihood favors controller parameters whose rollouts more closely reproduce the demonstrated lateral motion and steering.

{\emergencystretch=3em%
The kinematic bicycle model allows the controller parameters to describe the driver’s preferred path and look-ahead distance in geometric terms.
Vehicle-specific characteristics are handled separately at deployment through compensation for understeer and sideslip~\cite{seo,kapania} (Section~\ref{sec:hil-setup}), allowing the personalized parameters to be retained while the compensation is adapted to the vehicle.
This formulation avoids requiring tire and inertial parameters in the inference model, which could be confounded with the controller parameters if jointly estimated, and keeps the many closed-loop rollouts required by MCMC computationally inexpensive.
The HIL evaluation in Section~\ref{sec:held-out} assesses whether the inferred behavior is preserved with vehicle dynamics and a physical steering wheel.\par}

\subsection{Safety-aware inference target}\label{sec:safety}

The joint prior over \(\theta\) is
\begin{equation}
  \label{eq:prior}
p(\theta)\propto p(a)p(b)p(\varphi)\mathcal{N}(\symbfup{f};\symbfup{m},K_{\varphi})\indicator_{\mathcal{Y}}(a,b)
\end{equation}
where \(\mathcal{Y}\) is the set of \((a,b)\) satisfying \(|a+b\kappa|\leq y_{\max}\) for all calibration curvatures, and \(\indicator_{\mathcal{Y}}\) is its indicator function.
The Gaussian density \(\mathcal{N}(\symbfup{f};\symbfup{m},K_{\varphi})\) is the GP prior of Eq.~\eqref{eq:gp-prior}.

The trajectory likelihood in Eq.~\eqref{eq:likelihood} measures only how closely the rollouts reproduce the demonstrations and does not account for safety margins.
Demonstrations that violate predefined safety limits are therefore excluded from calibration. Rollouts that enter a warning region near these limits are penalized through a safety cost.

The safety cost is defined on lane clearance and guides the inference toward parameter regions that produce safer rollouts.
It remains zero until the warning threshold is crossed and then increases quadratically as the rollout moves closer to the lane boundary.
For rollout condition \(j\), let \(c_{j}\) denote the minimum clearance between the vehicle body and the lane boundaries. The safety cost is
\begin{equation}
  \label{eq:safety-cost}
J_{\mathrm{safety}}=\sum_{j=1}^{N_{\mathcal{S}}} {\left[\frac{c_{0}-c_{j}}{c_{0}}\right]}_{+}^{2}
\end{equation}
where \([x]_{+}=\max (0,x)\), \(c_{0}\) is the warning threshold and \(N_{\mathcal{S}}\) is the number of rollout conditions.
The cost is normalized by \(c_{0}\) so that each condition adds one to \(J_{\mathrm{safety}}\) when its rollout reaches the lane boundary, regardless of the chosen threshold. Other costs computed from the rollout, such as a yaw rate cost against the tire–road friction limit or a lateral jerk cost, can be added to \(J_{\mathrm{safety}}\) without changing the sampling procedure.

The resulting safety-aware inference target is
\begin{equation}
  \label{eq:target}
\begin{aligned}
  &q_{\lambda}(\theta \mid D_{\mathrm{cal}},\mathcal{S}) \\
  &\propto p(D_{\mathrm{cal}}\mid \theta,\mathcal{S})p(\theta)\exp (-\lambda J_{\mathrm{safety}})
\end{aligned}
\end{equation}
The weight \(\lambda\) scales the safety penalty relative to the trajectory likelihood.
Parameter values whose rollouts reproduce the demonstrations but enter the warning region therefore receive less probability mass than under the likelihood alone.

\subsection{MCMC sampling}\label{sec:sampling}

We sample the safety-aware target in Eq.~\eqref{eq:target} using a three-block Metropolis-within-Gibbs scheme. Both the trajectory likelihood and the safety cost require a closed-loop rollout, whereas the prior does not. We therefore write the terms that depend on the rollout as follows:
\begin{equation}
  \label{eq:rollout-term}
T_{\lambda}(\theta)=\log p(D_{\mathrm{cal}}\mid \theta,\mathcal{S})-\lambda J_{\mathrm{safety}}(\theta)
\end{equation}
so that the target of Eq.~\eqref{eq:target} reads
\begin{equation}
  \label{eq:factorization}
q_{\lambda}\propto \exp (T_{\lambda}(\theta))p(\theta)
\end{equation}
Proposals whose preferred-path parameters lie outside \(\mathcal{Y}\) have zero prior density under Eq.~\eqref{eq:prior} and are rejected before \(T_{\lambda}\) is evaluated.

The term \(T_{\lambda}\) is not differentiable everywhere, so we use random-walk Metropolis proposals that do not require gradients.
The blocks follow the dependencies between the variables and the separation of the target into rollout-dependent and prior factors shown in Eq.~\eqref{eq:factorization}. Blocks A and C modify the latent function and therefore require new closed-loop rollouts, whereas Block B updates only the GP hyperparameters and reuses the current rollouts.
Hyperparameters are sampled in logarithmic coordinates, \(\symbfup{\eta}=(\log \sigma_{f},\log \ell)\), with the prior density transformed accordingly.

Block A updates the preferred-path parameters and latent function jointly while holding the GP hyperparameters fixed, because they together determine the closed-loop trajectory.
The latent function is represented through a fixed reference transformation,
\begin{equation}
\symbfup{f}=\symbfup{m}+B_{\mathrm{ref}}\symbfup{\xi}_{\mathrm{ref}}
\end{equation}
where \(B_{\mathrm{ref}}{B_{\mathrm{ref}}}^{T}=K_{\mathrm{ref}}\) is the covariance matrix from Eq.~\eqref{eq:kernel} evaluated at fixed reference hyperparameters, and \(\symbfup{\xi}_{\mathrm{ref}}\) denotes the corresponding latent coordinates. A symmetric Gaussian random walk proposes changes to \((a,b,\symbfup{\xi}_{\mathrm{ref}})\).
Acceptance depends on the trajectory likelihood, safety penalty, preferred-path priors, and GP prior evaluated using the current hyperparameters.

Block B updates the GP hyperparameters while holding \((a,b,\symbfup{f})\) fixed, so the rollouts and \(T_{\lambda}\) are unchanged.
The acceptance probability therefore depends only on the hyperparameter prior and the GP density at the current latent function, which measures how plausible that function is under the proposed hyperparameters.

Block B alone changes the hyperparameters only slowly because the latent function constrains them strongly. Block C therefore updates the GP hyperparameters and latent function jointly to improve their exploration while \((a,b)\) remain fixed.
With \(B_{\varphi}{B_{\varphi}}^{T}=K_{\varphi}\), the current standardized latent coordinates are computed as
\begin{subequations}
\begin{equation}
  \label{eq:white-coordinates}
\symbfup{\nu}={B_{\varphi}}^{-1}(\symbfup{f}-\symbfup{m})
\end{equation}
A symmetric proposal for \(\symbfup{\eta}^{\prime}\) then determines
\begin{equation}
  \label{eq:coupled-function}
\symbfup{f}^{\prime}=\symbfup{m}+B_{\varphi^{\prime}}\symbfup{\nu}
\end{equation}
\end{subequations}

Preserving \(\symbfup{\nu}\) allows the latent function to change consistently with the covariance of the proposed hyperparameters~\cite{murray}. Under the GP prior, \(\symbfup{\nu}\) is standard normal for any \(\varphi\), so the GP-prior ratio cancels with the Jacobian of the transformation. The acceptance probability therefore involves only the trajectory likelihood, safety penalty, and hyperparameter prior.
With \({T_{\lambda}}^{\prime}=T_{\lambda}(\theta^{\prime})\), the coupled update is accepted with probability
\begin{equation}
  \label{eq:acceptance}
\min \left\{1,\exp ({T_{\lambda}}^{\prime}-T_{\lambda})\frac{p_{\eta}(\symbfup{\eta}^{\prime})}{p_{\eta}(\symbfup{\eta})}\right\}
\end{equation}
where \(p_{\eta}\) is the hyperparameter prior in log coordinates.

Each block accepts or rejects its proposed variables jointly, and Blocks A–C are applied using the state from the preceding block. Algorithm~\ref{alg:sweep} summarizes one sweep.

During warmup, the proposal covariance and scale of each block are adapted from the states of all chains at the end of each sweep, so all chains use the same proposal covariance for each block.
For each block \(r\), the proposal covariance is updated as
\begin{equation}
  \label{eq:proposal-covariance}
\begin{aligned}
  &Q_{r}=s_{r}\frac{{2.38}^{2}}{d_{r}} \\
  &\times [(1-\gamma)\widehat{C}_{r}+\gamma \operatorname{diag}(\operatorname{diag}(\widehat{C}_{r}))+\varepsilon_{0}D_{0,r}]
\end{aligned}
\end{equation}
where \(\widehat{C}_{r}\) is the sample covariance of the block variables collected from all chains over the current adaptation interval (for Block C, the covariance estimated for Block B is used because both blocks propose \(\symbfup{\eta}\)), \(d_{r}\) is the block dimension, \(D_{0,r}\) is diagonal with the squared initial proposal scales, \(\gamma\) and \(\varepsilon_{0}\) are small shrinkage and ridge weights, and \(s_{r}\) is a scalar tuned from the recent acceptance rate.
This construction follows adaptive Metropolis~\cite{haario} and aligns each proposal with the correlations among the variables of its block.
All adaptation is frozen before retaining samples.

\begin{algorithm}[!t]
\caption{One Metropolis-within-Gibbs sweep with a coupled GP update}
\label{alg:sweep}
\small\raggedright
\begin{algorithmic}[1]
\Require Current state \((a,b,\symbfup{\xi}_{\mathrm{ref}},\symbfup{\eta})\), cached \(T_{\lambda}(\theta)\), fixed \((\symbfup{m},B_{\mathrm{ref}})\), proposal covariances \(Q_{\mathrm{A}},Q_{\mathrm{B}},Q_{\mathrm{C}}\), and wide-proposal settings \(p_{\mathrm{w}},\sigma_{\mathrm{w}}\). Variables not proposed in a block remain unchanged, and infeasible candidates are rejected before rollout.
\State \(\symbfup{f}\gets\symbfup{m}+B_{\mathrm{ref}}\symbfup{\xi}_{\mathrm{ref}}\), \(\varphi\gets\exp(\symbfup{\eta})\), \(\symbfup{x}\gets{(a,b,\symbfup{\xi}_{\mathrm{ref}}^{T})}^{T}\)
\Statex \textbf{Block A}: preferred-path parameters and latent function, with \(\symbfup{\eta}\) fixed
\State With probability \(1-p_{\mathrm{w}}\), draw \(\symbfup{x}^{\prime}\sim\mathcal{N}(\symbfup{x},Q_{\mathrm{A}})\). Otherwise, draw \(j\) uniformly from \(\{1,\ldots,M\}\) and change only \(\xi_{\mathrm{ref},j}^{\prime}=\xi_{\mathrm{ref},j}+\sigma_{\mathrm{w}}\varepsilon\), \(\varepsilon\sim\mathcal{N}(0,1)\)
\State \(\symbfup{f}^{\prime}\gets\symbfup{m}+B_{\mathrm{ref}}\symbfup{\xi}_{\mathrm{ref}}^{\prime}\), \(l_{\mathrm{d}}^{\prime}(v_{i})\gets\exp(f_{i}^{\prime})\), and evaluate \(T_{\lambda}^{\prime}=T_{\lambda}(\theta^{\prime})\) by rollout
\State Accept \(\symbfup{x}^{\prime}\) with probability \({\min\{1,\exp(T_{\lambda}^{\prime}-T_{\lambda})\,p(\theta^{\prime})/p(\theta)\}}\)
\Statex \textbf{Block B}: GP hyperparameters, with \((a,b,\symbfup{f})\) fixed
\State Draw \(\symbfup{\eta}^{\prime}\sim\mathcal{N}(\symbfup{\eta},Q_{\mathrm{B}})\), \(\varphi^{\prime}\gets\exp(\symbfup{\eta}^{\prime})\)
\State Reuse the cached rollout evaluation and accept \(\symbfup{\eta}^{\prime}\) with probability \({\min\bigl\{1,\mathcal{N}(\symbfup{f};\symbfup{m},K_{\varphi^{\prime}})\,p_{\eta}(\symbfup{\eta}^{\prime})/[\mathcal{N}(\symbfup{f};\symbfup{m},K_{\varphi})\,p_{\eta}(\symbfup{\eta})]\bigr\}}\)
\Statex \textbf{Block C}: coupled GP hyperparameters and latent function, with \((a,b)\) fixed
\State \(\symbfup{\nu}\gets B_{\varphi}^{-1}(\symbfup{f}-\symbfup{m})\), draw \(\symbfup{\eta}^{\prime}\sim\mathcal{N}(\symbfup{\eta},Q_{\mathrm{C}})\), \(\varphi^{\prime}\gets\exp(\symbfup{\eta}^{\prime})\)
\State \(\symbfup{f}^{\prime}\gets\symbfup{m}+B_{\varphi^{\prime}}\symbfup{\nu}\), \(l_{\mathrm{d}}^{\prime}(v_{i})\gets\exp(f_{i}^{\prime})\), and evaluate \(T_{\lambda}^{\prime}\) by rollout
\State Accept \((\symbfup{\eta}^{\prime},\symbfup{f}^{\prime})\) jointly with probability \({\min\{1,\exp(T_{\lambda}^{\prime}-T_{\lambda})\,p_{\eta}(\symbfup{\eta}^{\prime})/p_{\eta}(\symbfup{\eta})\}}\)
\State Synchronize the reference coordinates, \(\symbfup{\xi}_{\mathrm{ref}}\gets B_{\mathrm{ref}}^{-1}(\symbfup{f}-\symbfup{m})\)
\If{warmup}
\State Store the end-of-sweep state, including repeats after rejection. At scheduled checkpoints, update \(Q_{\mathrm{A}}\), \(Q_{\mathrm{B}}\) and \(Q_{\mathrm{C}}\) by Eq.~\eqref{eq:proposal-covariance} from all chains together
\Else
\State Keep \(Q_{\mathrm{A}},Q_{\mathrm{B}},Q_{\mathrm{C}}\) and all other proposal parameters fixed, and retain the current joint state \(\theta\), including repeats
\EndIf
\end{algorithmic}
\end{algorithm}

\subsection{From posterior to deployed controller}\label{sec:deployment}

MCMC yields retained samples \(\theta^{(i)}\), \(i=1,\ldots,N\), whereas the deployed controller needs a single preferred-offset law and one look-ahead function. We therefore use posterior expectations estimated from the retained samples using Eq.~\eqref{eq:expectation}.

The Monte Carlo estimates of the posterior expectations of \(a\) and \(b\) are \(\overline{a}=\frac{1}{N}\sum_{i=1}^{N} a^{(i)}\) and \(\overline{b}=\frac{1}{N}\sum_{i=1}^{N} b^{(i)}\), respectively.
Since Eq.~\eqref{eq:preferred-offset} is affine in \(a\) and \(b\), the deployed preferred offset is
\begin{equation}
  \label{eq:deployed-offset}
\overline{y}_{\mathrm{off}}(\kappa)=\mathbb{E}_{q_{\lambda}}[a+b\kappa]\approx \overline{a}+\overline{b}\kappa .
\end{equation}

For the look-ahead, each retained sample defines a conditional GP prediction~\cite{rasmussen} at a given vehicle speed \(v\),
\begin{equation}
f_{*}(v)\mid \symbfup{f}^{(i)},\varphi^{(i)}\sim \mathcal{N}(\mu_{*}^{(i)}(v),\sigma_{*}^{2(i)}(v)),
\end{equation}
where
\begin{equation}
\begin{aligned}
  &\mu_{*}^{(i)}(v)=m(v)+\symbfup{k}_{*}^{(i)\mathrm{T}}K_{\varphi^{(i)}}^{-1}(\symbfup{f}^{(i)}-\symbfup{m}), \\
  &\sigma_{*}^{2(i)}(v)=k_{**}^{(i)}-\symbfup{k}_{*}^{(i)\mathrm{T}}K_{\varphi^{(i)}}^{-1}\symbfup{k}_{*}^{(i)}.
\end{aligned}
\end{equation}
Here, \(\symbfup{k}_{*}^{(i)}\) contains the covariances between the given speed \(v\) and the calibration speeds, and \(k_{**}^{(i)}\) is the prior variance at \(v\) under \(\varphi^{(i)}\).
Since \(l_{\mathrm{d}}(v)=\exp (f_{*}(v))\), its conditional expectation for posterior draw \(i\) is
\begin{equation}
\begin{aligned}
  &\mathbb{E}[l_{\mathrm{d}}(v)\mid \symbfup{f}^{(i)},\varphi^{(i)}] \\
  &=\exp \left[\mu_{*}^{(i)}(v)+\frac{1}{2}\sigma_{*}^{2(i)}(v)\right].
\end{aligned}
\end{equation}

Averaging these conditional expectations over the retained draws yields a Monte Carlo estimate of the deployed look-ahead distance at speed \(v\),
\begin{equation}
  \label{eq:deployed-lookahead}
\begin{aligned}
  &\overline{l}_{\mathrm{d}}(v)=\mathbb{E}_{q_{\lambda}}[\mathbb{E}[l_{\mathrm{d}}(v)\mid \symbfup{f},\varphi]] \\
  &\approx \frac{1}{N}\sum_{i=1}^{N} \exp \left[\mu_{*}^{(i)}(v)+\frac{1}{2}\sigma_{*}^{2(i)}(v)\right].
\end{aligned}
\end{equation}
Because \(\varphi\) is averaged over its samples rather than fixed at a point estimate, the posterior dependence between \(\varphi\) and \(\symbfup{f}\) is preserved in the prediction.

The uncertainty of the look-ahead at a given speed arises from both the GP and the MCMC posterior, as the predictive variance of \(f_{*}(v)\) decomposes into
\begin{equation}
  \label{eq:predictive-variance}
\operatorname{Var}_{q_{\lambda}}[f_{*}(v)]=\mathbb{E}_{q_{\lambda}}[\sigma_{*}^{2}(v)]+\operatorname{Var}_{q_{\lambda}}[\mu_{*}(v)],
\end{equation}
where the first term is the expected conditional GP variance and the second is the variance of the conditional mean across the MCMC posterior samples. The conditional GP variance tends to be larger at speeds that are weakly supported by the calibration data.

{\emergencystretch=3em%
Eqs.~\eqref{eq:deployed-offset} and~\eqref{eq:deployed-lookahead} define the deployed controller through posterior expectations, so its dependence on the safety weight \(\lambda\) can be characterized using Eq.~\eqref{eq:target}.
To examine this dependence, we collect the preferred-offset coefficients \(a\) and \(b\), together with the mean look-ahead distance of the conditional GP prediction at a given speed \(v\), into a vector for each posterior sample,
\begin{equation}
  \label{eq:deployed-vector}
\symbfup{g}(\theta)={\left[a,\;b,\;\mathbb{E}[l_{\mathrm{d}}(v)\mid \symbfup{f},\varphi]\right]}^{T},
\end{equation}
so that \(\overline{\symbfup{g}}_{\lambda}=\mathbb{E}_{q_{\lambda}}[\symbfup{g}(\theta)]={[\overline{a},\overline{b},\overline{l}_{\mathrm{d}}(v)]}^{T}\) contains the deployed parameters.
Let \(q_{0}\) denote the target of Eq.~\eqref{eq:target} with \(\lambda=0\), so that \(q_{\lambda}\propto q_{0}w_{\lambda}\) with the penalty factor \(w_{\lambda}=\exp (-\lambda J_{\mathrm{safety}})\). The deployed parameters then differ from their values without the penalty by
\begin{equation}
  \label{eq:lambda-change}
\overline{\symbfup{g}}_{\lambda}-\overline{\symbfup{g}}_{0}=\frac{\operatorname{Cov}_{q_{0}}[\symbfup{g}(\theta),w_{\lambda}(\theta)]}{\mathbb{E}_{q_{0}}[w_{\lambda}(\theta)]}.
\end{equation}\par}

By the Cauchy–Schwarz inequality, each component \(g_{k}\) of \(\symbfup{g}\) satisfies
\begin{equation}
  \label{eq:lambda-bound}
\left|\overline{g}_{\lambda,k}-\overline{g}_{0,k}\right|\leq \sqrt{\operatorname{Var}_{q_{0}}[g_{k}]}\,\frac{\sqrt{\operatorname{Var}_{q_{0}}[w_{\lambda}]}}{\mathbb{E}_{q_{0}}[w_{\lambda}]}.
\end{equation}

This bound shows that a parameter's posterior uncertainty without the safety penalty limits how much its deployed value can change.
Since the second factor is common to all components, parameters with greater posterior uncertainty allow larger changes, whereas parameters with small posterior uncertainty remain close to their values without the penalty.
The actual direction and magnitude of each change are determined by its covariance with \(w_{\lambda}\), a weight that is small where the safety cost is high.
The safety penalty thus acts as an uncertainty-dependent margin, changing each deployed parameter toward lower safety costs only as far as its posterior uncertainty allows, without a margin tuned by hand for each condition.

\section{Offline personalization evaluation}\label{sec:experiments}

\subsection{Driver data and inference settings}\label{sec:data}

\begin{figure*}[!tb]
  \centering
  \includegraphics[trim=0 7pt 0 0,clip]{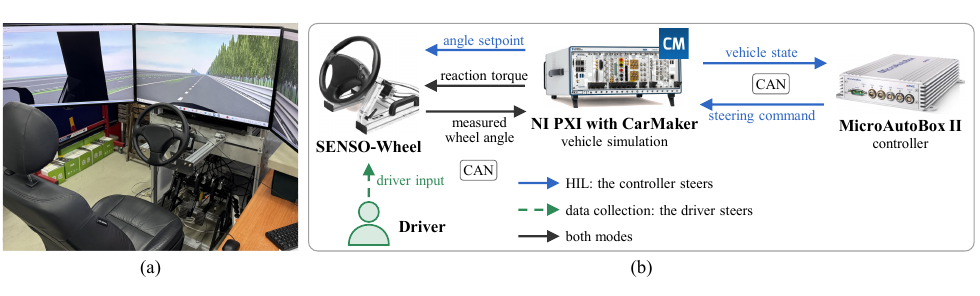}
  \caption{(a) Driving simulator used for data collection and HIL evaluation. (b) Devices and signals in the two operating modes.}
  \label{fig:driving-simulator}
\end{figure*}

Demonstrations were recorded from two drivers on the driving simulator shown in Fig.~\ref{fig:driving-simulator}, which also serves for the HIL evaluation in Section~\ref{sec:held-out}. IPG CarMaker runs the vehicle and road models on an NI PXI real-time target, and NI VeriStand maps the I/O and controller area network (CAN) signals.
The SENSO-Wheel serves as the driver interface and is connected to the real-time target over CAN. During data collection, the drivers drove the simulated vehicle with the SENSO-Wheel and received a steering reaction torque computed by CarMaker.

Each road consisted of a straight approach followed by a constant-radius left curve.
Nineteen conditions were defined for this study, with speeds ranging from 30 to 70 km/h and lane-center curvatures from 0.0040 to 0.0098 \(\mathrm{m^{-1}}\) (road radii from 250 to 100 m).
Sixteen were used for calibration, and the three held-out conditions were used for the comparison in Section~\ref{sec:held-out}.

Rollouts used \(\Delta t=0.05\,\mathrm{s}\) and \(L=2.97\,\mathrm{m}\) on a 3.5 m lane, and the offset transition covered the 20 m before curve entry.
The likelihood compared each demonstration with its rollout every 0.3 s. To avoid overly precise inference from serially correlated residuals, the likelihood was scaled by an effective sample size~\cite{ribatet,brownhund}, that is, the error variances estimated from the demonstrations were inflated by the residual autocorrelation time, giving standard deviations of 0.40 m, 0.029 rad, and 0.0078 rad for lateral offset, heading error, and steering angle.
The safety cost used \(c_{0}=0.25\,\mathrm{m}\), and the safety penalty used \(\lambda=225\).

The priors were \(a\sim \mathcal{N}(0,{0.3}^{2})\), \(b\sim \mathcal{N}(0,{40}^{2})\), \(\sigma_{f}\sim \operatorname{HalfNormal}(0.5)\),
and \(\log \ell\sim \mathcal{N}(0,{0.7}^{2})\), with the offset bound \(y_{\max}=0.6\,\mathrm{m}\).
The GP prior used \(m(v)=\log(1\,\mathrm{s}\cdot v)\), with standard deviations of 0.5 and 1 for the level and slope. A total of 64 chains were started from four prespecified states, 16 chains per state.
Each chain ran 50,000 warmup sweeps and 1,100,000 retained sweeps, and the retained sequence was analyzed without thinning. MCMC sampling was performed offline using a single NVIDIA GeForce RTX 4060 GPU.

Sampling was assessed for all 13 parameters in physical coordinates using rank-normalized split-\(\widehat{R}\) with folding, bulk effective sample size (ESS), and tail ESS at the 5th and 95th percentiles~\cite{vehtari}.
The recommended values \(\widehat{R}<1.01\) and ESS > 400 serve as references, the former being stricter than the 1.1 threshold of earlier guidance~\cite{gelman}.

\subsection{Synthetic recovery}\label{sec:recovery}

\begin{figure}[!b]
  \centering
  \includegraphics[width=\linewidth]{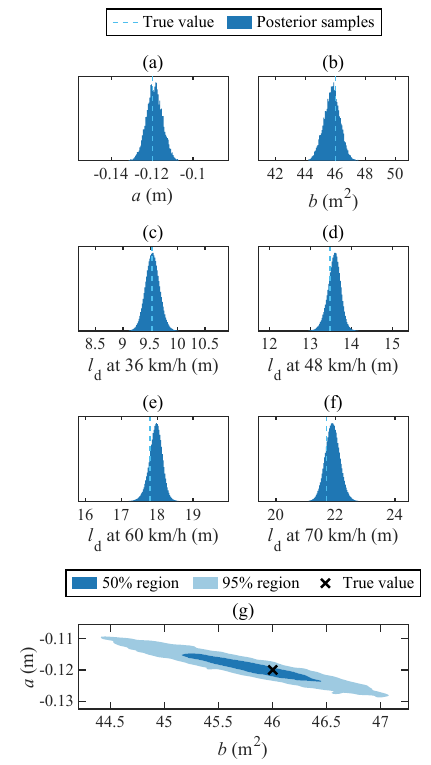}
  \caption{Synthetic recovery of a known controller. (a), (b)~Preferred-offset parameters. (c)–(f)~Look-ahead distance. (g)~Joint posterior of the preferred-offset parameters.}
  \label{fig:synthetic-recovery}
\end{figure}

To check whether the inference recovers a known controller within its 95\% intervals, the procedure was applied to synthetic data generated by the controller with \(a=-0.12\,\mathrm{m}\), \(b=46\,\mathrm{m}^{2}\), and \(l_{\mathrm{d}}=0.78v+0.0173v^{2}\) (\(v\) in m/s), chosen to be close to the driver parameter estimates reported in Section~\ref{sec:calibration}.
This controller was rolled out on the 16 calibration conditions, and independent Gaussian noise at sensor level, with standard deviations of 0.037 m, 0.0031 rad, and 0.0011 rad, was added, with the same values used as the likelihood standard deviations.
We used the priors and sampler described in Section~\ref{sec:data} and set \(\lambda=0\) to disable the safety penalty. Without the penalty, the target is the posterior of the model that generated the data, so this check tests recovery alone.

The \(\widehat{R}\) values of \(a\) and \(b\) were 1.018 and 1.016, above the reference but below the earlier threshold of 1.1, and those of the look-ahead distance at the nine speeds ranged from 1.000 to 1.001.
The true values fell within the 95\% intervals for all 14 quantities assessed: \(a\), \(b\), the look-ahead distance at the nine speeds, and the preferred offset at three curvatures spanning the calibration range (see Fig.~\ref{fig:synthetic-recovery}(a)–(f) for selected marginal posteriors).
Relative to the truth, the posterior means of \(a\) and \(b\) differed by +1.1\% and −0.6\%, those of the look-ahead distance at the nine speeds by −0.9\% to +0.9\%, and those of the preferred offset by at most 1.3 mm.
The errors in \(a\) and \(b\) lay along the ridge of their joint posterior, a direction that the calibration curvatures constrain only weakly, with the true \((a,b)\) inside the 50\% region (Fig.~\ref{fig:synthetic-recovery}(g)). The chains also mixed slowly along this ridge, which explains the \(\widehat{R}\) values of \(a\) and \(b\), although their bulk ESS values of 4,230 and 4,550 were well above 400.

\subsection{Calibration results}\label{sec:calibration}

We inferred the personalized controller from the 16 calibration trajectories of one driver, using the settings of Section~\ref{sec:data}. We first check that the chains converged and then examine whether the posterior identifies a curvature-dependent preferred offset and a speed-dependent look-ahead distance.

\subsubsection{Sampling diagnostics}\label{sec:diagnostics}

\begin{table}[!tb]
  \centering\small
  \caption{MCMC convergence diagnostics}
  \label{tab:diagnostics}
  \begin{tabular*}{\linewidth}{@{\extracolsep{\fill}}lrrr@{}}
    \toprule
    \textbf{Parameter} & \(\widehat{\symbfit{R}}\) & \textbf{Bulk ESS} & \textbf{Tail ESS} \\
    \midrule
    \(a\) & 1.008 & 7,834 & 15,414 \\
    \(b\) & 1.008 & 7,759 & 15,928 \\
    \(l_{\mathrm{d}}\) (9 speeds) & 1.002 & 45,541 & 70,975 \\
    \(\sigma_{f}\) & 1.000 & 290,865 & 668,392 \\
    \(\ell\) & 1.001 & 112,740 & 143,309 \\
    \bottomrule
  \end{tabular*}\par
  \smallskip
  {\footnotesize\raggedright For \(l_{\mathrm{d}}\), the table reports the maximum \(\widehat{R}\) and minimum ESS values across the nine speeds.\par}
\end{table}

All parameters satisfied both references (Table~\ref{tab:diagnostics}). The parameters \(a\) and \(b\) mixed the slowest, and their \(\widehat{R}\) fell below the earlier threshold of 1.1 after about 100,000 retained sweeps per chain.
The Monte Carlo standard errors of \(\overline{a}\), \(\overline{b}\), and \(\overline{l}_{\mathrm{d}}\) were below 0.2\% of the deployed values, so the deployed controller is insensitive to the particular sampling run. After the first 200,000 sweeps (30 h), \(\overline{a}\), \(\overline{b}\), and \(\overline{l}_{\mathrm{d}}\) at the nine speeds differed by less than 0.7\% from their values after all 1,150,000 sweeps (179 h).

\subsubsection{Inferred preferred path and look-ahead}\label{sec:inferred-controller}

\begin{table}[!tb]
  \centering\small
  \caption{Posterior means and 95\% intervals (2.5th to 97.5th percentiles)}
  \label{tab:posterior}
  \begin{tabular*}{\linewidth}{@{\extracolsep{\fill}}lrr@{}}
    \toprule
    \textbf{Parameter} & \textbf{Mean} & \textbf{95\% interval} \\
    \midrule
    \(a\) (m) & −0.1167 & [−0.1955, −0.0372] \\
    \(b\) (m²) & 46.08 & [35.27, 56.87] \\
    \(l_{\mathrm{d}}\) (m) & & \\
    \quad 30 km/h & 7.24 & [5.64, 8.93] \\
    \quad 36 km/h & 9.33 & [7.86, 10.88] \\
    \quad 42 km/h & 11.57 & [10.03, 13.33] \\
    \quad 45 km/h & 12.73 & [11.08, 14.80] \\
    \quad 48 km/h & 13.88 & [12.05, 16.32] \\
    \quad 54 km/h & 16.01 & [13.85, 18.80] \\
    \quad 60 km/h & 17.93 & [15.72, 20.11] \\
    \quad 66 km/h & 19.79 & [17.56, 21.30] \\
    \quad 70 km/h & 21.17 & [18.33, 23.01] \\
    \(\sigma_{f}\) & 0.1792 & [0.0053, 0.6488] \\
    \(\ell\) & 1.724 & [0.341, 4.909] \\
    \bottomrule
  \end{tabular*}
\end{table}

\begin{figure}[!tb]
  \centering
  \includegraphics[width=\linewidth]{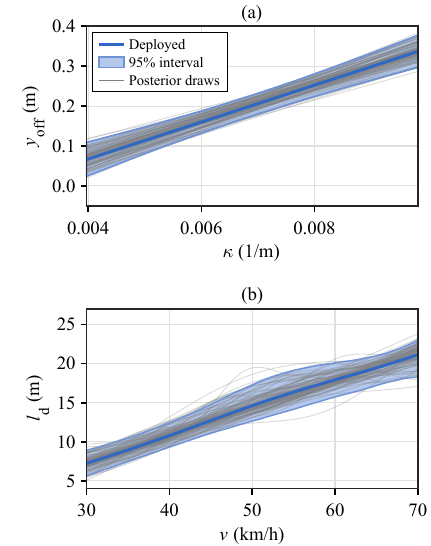}
  \caption{Inferred preferred offset (a) and look-ahead distance (b), with 100 posterior draws (grey).}
  \label{fig:inferred-controller}
\end{figure}

The slope \(b\) is positive over its whole 95\% interval (Table~\ref{tab:posterior}), so the preferred offset moves toward the inside of the curve as curvature increases.
The parameters \(a\) and \(b\) trade off along a ridge, so their individual intervals are broad.
Along this ridge, the preferred offset \(a+b\kappa\) remains nearly unchanged over the calibration curvature range, so the offset is more tightly constrained than \(a\) and \(b\) individually. At each curvature in the calibration range, its 95\% interval was 0.05–0.08 m wide, compared with 0.16 m for \(a\) alone.
The deployed preferred offset \(\overline{a}+\overline{b}\kappa\) increased from 0.07 m at the smallest calibration curvature to 0.34 m at the largest (Fig.~\ref{fig:inferred-controller}(a)).

For this driver, the posterior supports an increase in look-ahead distance with speed, while its local rate remains uncertain (Fig.~\ref{fig:inferred-controller}(b)).

\subsection{Personalization across drivers}\label{sec:drivers}

\begin{figure}[!t]
  \centering
  \includegraphics[width=\linewidth]{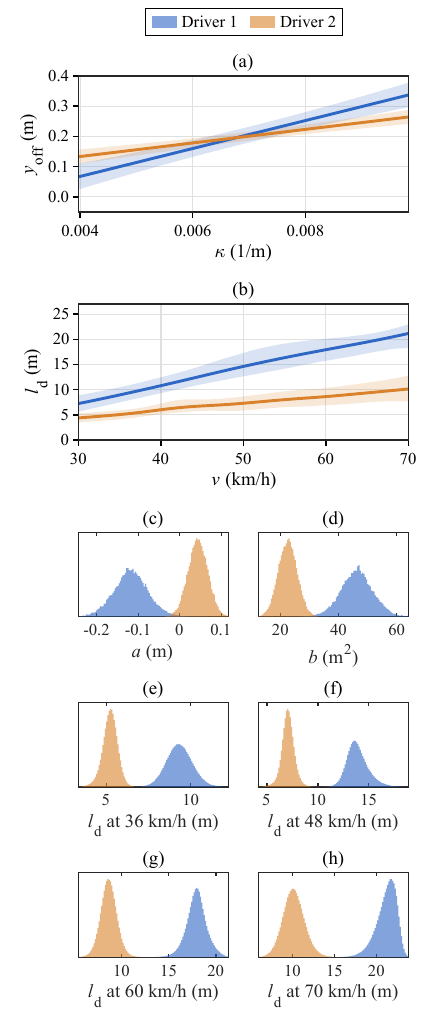}
  \caption{Preferred offset (a) and look-ahead distance (b) for two drivers, with 95\% posterior intervals (shaded). Panels (c)–(h) show the marginal posteriors.}
  \label{fig:drivers}
\end{figure}

\begin{table}[!b]
  \centering\small
  \caption{Posterior means and 95\% intervals (2.5th to 97.5th percentiles) of driver~2}
  \label{tab:posterior-driver2}
  \begin{tabular*}{\linewidth}{@{\extracolsep{\fill}}lrr@{}}
    \toprule
    \textbf{Parameter} & \textbf{Mean} & \textbf{95\% interval} \\
    \midrule
    \(a\) (m) & 0.0445 & [−0.0003, 0.0897] \\
    \(b\) (m²) & 22.33 & [16.14, 28.46] \\
    \(l_{\mathrm{d}}\) (m) & & \\
    \quad 30 km/h & 4.39 & [3.48, 5.35] \\
    \quad 36 km/h & 5.22 & [4.28, 6.05] \\
    \quad 42 km/h & 6.40 & [5.47, 7.91] \\
    \quad 45 km/h & 6.77 & [5.80, 8.05] \\
    \quad 48 km/h & 7.05 & [5.73, 8.32] \\
    \quad 54 km/h & 7.89 & [6.43, 9.46] \\
    \quad 60 km/h & 8.63 & [6.92, 10.37] \\
    \quad 66 km/h & 9.50 & [7.56, 11.58] \\
    \quad 70 km/h & 10.11 & [7.66, 12.77] \\
    \(\sigma_{f}\) & 0.2027 & [0.0068, 0.7027] \\
    \(\ell\) & 1.541 & [0.248, 4.863] \\
    \bottomrule
  \end{tabular*}
\end{table}

To examine whether the method can distinguish between drivers, we applied the same inference framework to a second driver (driver~2), who completed the same 16 calibration conditions as the driver of Section~\ref{sec:calibration} (driver~1). The model, priors, safety penalty, and sampling procedure were unchanged. The likelihood standard deviations were estimated separately from each driver's demonstrations. For driver~2, they were 0.23 m, 0.015 rad, and 0.0063 rad for lateral offset, heading error, and steering angle, respectively. As for driver~1, all parameters of driver~2 satisfied both references, with bulk ESS of at least 9,225 and \(\widehat{R}\) of 1.007 for \(a\) and \(b\). Fig.~\ref{fig:drivers} compares the inferred controller functions and selected marginal posterior distributions, while Tables~\ref{tab:posterior} and \ref{tab:posterior-driver2} report their posterior summaries.

The inferred preferred-offset functions crossed within the calibration curvature range and differed by up to 0.07~m at its ends (Fig.~\ref{fig:drivers}(a)). Driver~2 had a smaller slope \(b\) than driver~1, with nonoverlapping 95\% intervals (Fig.~\ref{fig:drivers}(d), Tables~\ref{tab:posterior} and \ref{tab:posterior-driver2}), indicating different inferred curvature dependencies rather than a constant lateral shift.

Driver~2's inferred look-ahead distance was shorter at every calibration speed and increased less with speed (Fig.~\ref{fig:drivers}(b), (e)–(h)). The 95\% intervals of the two drivers did not overlap at any calibration speed (Tables~\ref{tab:posterior} and \ref{tab:posterior-driver2}). In the pure pursuit law, a shorter look-ahead steers toward a closer point on the preferred path, so driver~2 tracks the preferred path more responsively than driver~1. Thus, under the same calibration conditions, the inference yielded distinguishable curvature-dependent preferred paths and speed-dependent look-ahead functions for the two drivers.

\subsection{Effect of the safety penalty}\label{sec:ablation}

\subsubsection{Posterior with and without the penalty}\label{sec:ablation-posterior}

\begin{figure*}[!tb]
  \centering
  \includegraphics[width=\linewidth]{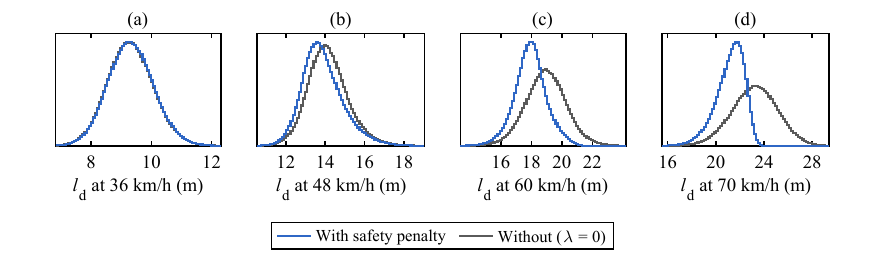}
  \caption{Marginal posteriors of the look-ahead distance at 36, 48, 60, and 70 km/h (a)–(d) with and without the safety penalty.}
  \label{fig:safety-posterior}
\end{figure*}

\begin{figure*}[!tb]
  \centering
  \includegraphics[width=\linewidth]{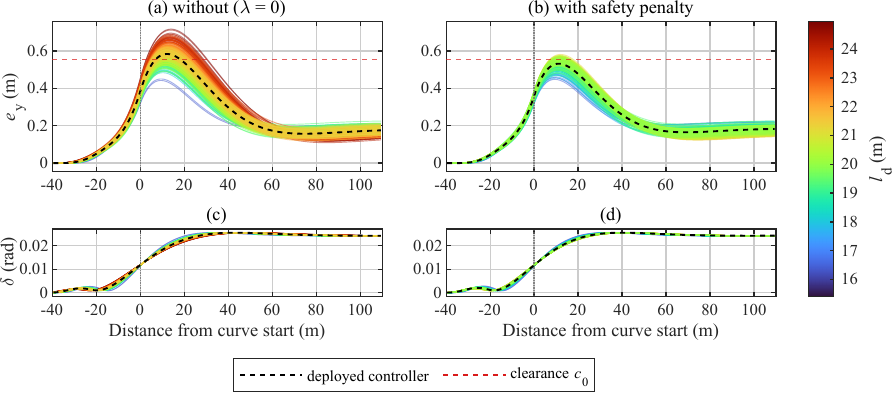}
  \caption{Rollouts of 256 posterior samples on the 120 m radius curve at 66 km/h.}
  \label{fig:safety-rollouts}
\end{figure*}

To isolate the effect of the safety penalty, inference was repeated with \(\lambda=0\) and all other settings unchanged.
The preferred offset changed by at most 4 mm.
The penalty had little effect on the deployed look-ahead at low speeds and shortened it at higher speeds (Fig.~\ref{fig:safety-posterior}).
Consistent with Eq.~\eqref{eq:lambda-bound}, the deployed look-ahead changed more at speeds where its 95\% interval without the penalty was wider.
At 70 km/h, the penalty suppressed the upper tail of the posterior, lowering the 97.5th percentile from 26.9 m to 23.0 m and leaving the 95\% interval extending 1.8 m above the deployed value and 2.8 m below it, whereas without the penalty the interval was nearly symmetric (Fig.~\ref{fig:safety-posterior}(d)).

A longer look-ahead makes the pure pursuit controller turn in earlier, so the vehicle passes closer to the inner lane boundary at curve entry. On the 120 m radius curve at 66 km/h (Fig.~\ref{fig:safety-rollouts}), the penalty suppressed the samples with the longest look-ahead, which reached the largest peak lateral offset at curve entry, while the steady-state lateral offset remained nearly the same.
The deployed controller with the penalty kept a minimum clearance above \(c_{0}=0.25\,\mathrm{m}\) on all 16 conditions. The penalty increased this clearance mainly at high speed. Without the penalty, the deployed controller entered the clearance warning region on two of these conditions (120 m radius at 66 km/h and 130 m radius at 70 km/h), with minimum clearances of 0.220 and 0.219 m, compared with 0.272 and 0.284 m with the penalty, respectively.
The penalty reduced the share of posterior samples whose rollout entered the clearance warning region on at least one of the 16 calibration conditions from 79.3\% to 29.1\%. With the penalty, the safety cost for each posterior sample implied a clearance of at least 0.18 m on all 16 calibration conditions. The penalty thus increased clearance for the deployed controller and across the posterior.

{\emergencystretch=3em%
The cost in trajectory fit was small, as the lateral root-mean-square (RMS) error of the deployed controller changed by at most 0.009 m per condition.
At every speed, the deployed look-ahead with the penalty remained within the 95\% interval of the posterior without it, so the penalty selects the safer part of the range supported by the demonstrations rather than overriding the driver's preference.\par}

\subsubsection{Comparison with deterministic identification}\label{sec:deterministic}

\begin{figure}[!tb]
  \centering
  \includegraphics[width=\linewidth,trim=0 6bp 0 0,clip]{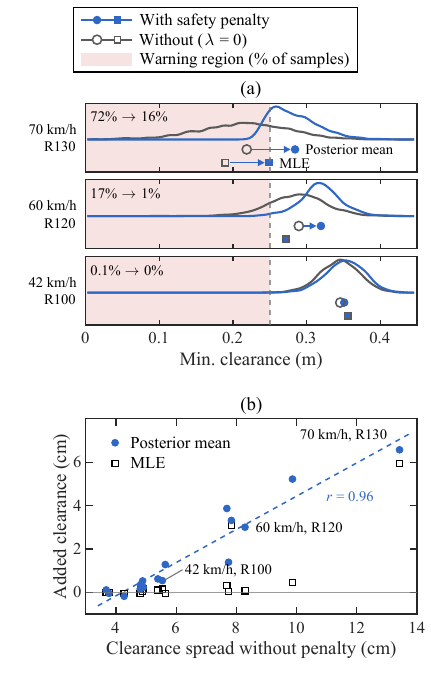}
  \caption{(a) Minimum clearance with and without the safety penalty. (b) Added clearance versus clearance spread without the penalty.}
  \label{fig:deterministic}
\end{figure}

We compared the proposed method with deterministic identification using the same rollout model, calibration data, and trajectory likelihood. The maximum-likelihood estimate (MLE) maximizes the rollout term of Eq.~\eqref{eq:rollout-term} without the prior,
\begin{equation}
  \label{eq:mle}
\hat{\theta}_{\lambda}=\operatorname*{arg\,max}_{a,b,\symbfup{f}}\,T_{\lambda}(\theta),
\end{equation}
which was computed by nonlinear least squares without and with the safety penalty (\(\lambda=0\) and \(\lambda=225\)), the latter yielding a penalized MLE. Eight starting points converged to the same solution.

The MLE and the deployed controller yielded nearly identical lateral RMS errors of 0.158–0.159 m over the calibration conditions, both without and with the safety penalty.
However, the safety penalty acts on the MLE only when its own rollout enters the warning region, whereas in the proposed method, it reweights the posterior according to the safety cost of each sample (Fig.~\ref{fig:deterministic}(a)).

On the 130 m radius curve at 70 km/h, the penalty increased the MLE clearance from 0.190 m to approximately \(c_0\), whereas the deployed controller's clearance rose well above \(c_0\). The distinction is more evident on the 120 m radius curve at 60 km/h. The MLE already satisfied the warning threshold with 0.272 m of clearance and changed little, yet 17\% of the posterior samples without the penalty entered the warning region. By reducing the posterior weight of such samples, the penalty increased the deployed controller's clearance by 3.0 cm, although the MLE itself satisfied the threshold. In contrast, on the 100 m radius curve at 42 km/h, only 0.1\% of the posterior samples without the penalty entered the warning region, and both the MLE and the deployed controller changed little. These results show that the safety penalty changes the deployed controller by accounting for safety costs across the posterior, without prescribing a separate clearance margin for each condition.

Fig.~\ref{fig:deterministic}(b) extends this comparison to all 16 conditions, with the clearance spread measured from the median to the 2.5th percentile of the posterior samples without the penalty.
For the deployed controller, added clearance increased with this spread (Pearson \(r=0.96\)). This trend is consistent with the uncertainty-dependent parameter changes described by Eq.~\eqref{eq:lambda-bound}. In contrast, the MLE showed substantial clearance changes only on the two curves at 70 km/h.

The proposed method therefore matched the trajectory fit of deterministic identification, while the safety penalty acted on its deployed controller as the uncertainty-dependent margin described in Section~\ref{sec:deployment}.

\section{Hardware-in-the-loop implementation and validation}\label{sec:held-out}

We evaluate whether the personalized controller, steering in place of the driver, reproduces the driver’s behavior in HIL with vehicle dynamics and a physical steering wheel. The deployed parameters of both drivers obtained in Sections~\ref{sec:calibration} and \ref{sec:drivers} are used without modification. After confirming that the deployed controller runs in real time, we verify that the physical wheel follows its command. We also verify that the resulting trajectories match their offline rollouts. We then compare the personalized and baseline controllers with the driver’s held-out recordings and evaluate lane clearance.

\subsection{HIL setup}\label{sec:hil-setup}

The deployed parameters \(\overline{a}\), \(\overline{b}\), and \(\overline{l}_{\mathrm d}(v)\) of Section~\ref{sec:deployment} were embedded in a pure pursuit controller on a dSPACE MicroAutoBox II. In the loop shown in Fig.~\ref{fig:driving-simulator}(b), CarMaker sent the position, yaw angle, and speed of the vehicle over CAN to the MicroAutoBox II. The MicroAutoBox II computed a steering-wheel angle command from the received vehicle state every 10~ms. The SENSO-Wheel tracked this command under position control, with gains tuned for stable tracking, adding a position-control torque to the steering reaction torque from CarMaker, and the measured wheel angle fed back into CarMaker as the steering input. Communication delays and the physical steering response therefore entered the closed loop. The vehicle and the road were simulated, while the controller hardware, the CAN bus, and the steering actuator were real.

The baseline pure pursuit controller used the same law with \(a=b=0\) and \(l_{\mathrm d}(v)=1\,\mathrm{s}\cdot v\). These are the prior means of \(a\) and \(b\) and the prior median of the look-ahead distance (Section~\ref{sec:data}), so the baseline represents the controller determined by the prior before any demonstration is observed.

Both controllers used the same compensation for the rear-axle sideslip and understeer of the vehicle~\cite{seo,kapania}. Assuming both effects are proportional to lateral acceleration, we used
\begin{equation}
\begin{aligned}
\kappa_{\mathrm c} &= \frac{\kappa_{\mathrm{PP}}}{1-2D_{\mathrm r}v^2/l_{\mathrm d}},\\
\delta_{\mathrm{cmd}} &= \arctan(L\kappa_{\mathrm c})+K_{\mathrm{us}}v^2\kappa_{\mathrm c},
\end{aligned}
\end{equation}
where \(\kappa_{\mathrm{PP}}=2\sin\alpha/l_{\mathrm d}\) is the pure pursuit curvature.
The coefficients \(D_{\mathrm r}=0.00402\) and \(K_{\mathrm{us}}=0.00197\), both in rad per \(\mathrm{m/s^2}\), describe the vehicle and were estimated from the calibration recordings.

The test conditions were the three held-out left turns of Section~\ref{sec:data}: H1 (190~m radius, 39~km/h), H2 (170~m, 55~km/h), and H3 (140~m, 65~km/h). Each controller was tested five times per condition.

\subsection{Execution in the HIL loop}\label{sec:hil-execution}

 At runtime, the controller evaluated the preferred offset \(\overline{a}+\overline{b}\kappa\) directly and interpolated \(\overline{l}_{\mathrm d}(v)\) from a table computed offline over a speed grid to avoid evaluating the posterior expectation at every control step. 
 Each control step on the MicroAutoBox II took less than 27~\textmu s, and no task overrun occurred.

\begin{figure}[!tb]
  \centering
  \includegraphics[width=\linewidth]{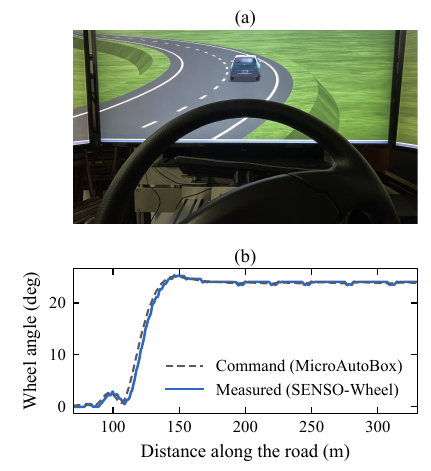}
  \caption{(a) Steering wheel and CarMaker view during an HIL run. (b) Commanded and measured steering-wheel angles of the personalized controller of driver~1 at H3.}
  \label{fig:wheel-tracking}
\end{figure}

With the tuned position-control gains, the measured wheel angle followed the command in all HIL runs with a delay of about 130~ms, and the RMS difference after this delay was about 0.2\(^{\circ}\) (Fig.~\ref{fig:wheel-tracking}). Most of the delay occurred between the MicroAutoBox II command and the wheel response in the hardware loop. This delay is comparable to that of an automated vehicle built on a production steering system, in which communication and mechanical clearance delayed the steering-wheel response to a command by about 200~ms~\cite{xu}. These results show that the controller obtained from offline inference runs within the 10~ms sample time and operates reliably with the physical steering wheel under reaction torque.

Next, we test how closely the HIL trajectories follow the offline rollouts predicted by the inference model. To separate the effect of the vehicle dynamics from that of the hardware loop, the same comparison is made for a CarMaker simulation with an ideal steering wheel, where the steering command is applied directly to the vehicle model.

In CarMaker with an ideal wheel, the RMS difference from the offline rollout was close to that in HIL at every condition (Table~\ref{tab:hil-comparison}). The physical steering loop therefore added little, and most of the difference arose between the inference model and the full vehicle simulation rather than in the hardware. Despite the steering delay of about 130~ms, the personalized HIL trajectories stayed close to their offline rollouts across all conditions, confirming that the controller inferred with the kinematic model transfers reliably to full vehicle dynamics and physical steering actuation.

\subsection{Driver reproduction and lane clearance}\label{sec:hil-reproduction}

For each driver, we compared both controllers with the same driver recordings using the RMS error in lateral offset \(e_y\), averaged over the five recordings and five HIL runs of each condition. We also evaluated the minimum clearance of each personalized controller over all repeats.

{\emergencystretch=3em%
For both drivers, the personalized controller reproduced the driver’s held-out trajectories more closely than the baseline in all three conditions, although the reduction in lateral RMS error was small for driver~1 at H2 (Table~\ref{tab:hil-comparison}). The personalized controllers maintained lane clearance above the 0.25~m warning threshold in every HIL run (Table~\ref{tab:hil-comparison}). These tests covered two drivers and three left turns within the calibration speed and curvature ranges.}\par

\begin{table}[!tbp]
  \centering\small\setlength{\tabcolsep}{3pt}
  \caption{Held-out errors and lane clearance by condition (m)}
  \label{tab:hil-comparison}
  \begin{tabular*}{\linewidth}{@{\extracolsep{\fill}}lrrrrrr@{}}
    \toprule
    & \multicolumn{3}{c}{\textbf{Driver 1}} & \multicolumn{3}{c@{}}{\textbf{Driver 2}} \\
    \cmidrule(lr){2-4}\cmidrule(l){5-7}
    & \textbf{H1} & \textbf{H2} & \textbf{H3} & \textbf{H1} & \textbf{H2} & \textbf{H3} \\
    \midrule
    \multicolumn{7}{@{}l}{RMS error against driver recordings} \\
    \quad Baseline HIL & 0.175 & 0.136 & 0.214 & 0.160 & 0.161 & 0.188 \\
    \quad Personalized HIL & 0.106 & 0.130 & 0.177 & 0.095 & 0.111 & 0.107 \\
    \multicolumn{7}{@{}l}{RMS difference from offline rollout, personalized controller} \\
    \quad CarMaker, ideal wheel & 0.013 & 0.034 & 0.055 & 0.008 & 0.017 & 0.026 \\
    \quad HIL & 0.013 & 0.038 & 0.064 & 0.008 & 0.018 & 0.030 \\
    \multicolumn{7}{@{}l}{Minimum clearance, personalized controller} \\
    \quad Offline & 0.63 & 0.52 & 0.37 & 0.63 & 0.60 & 0.55 \\
    \quad HIL & 0.65 & 0.56 & 0.45 & 0.63 & 0.61 & 0.57 \\
    \bottomrule
  \end{tabular*}
\end{table}

\section{Conclusion}\label{sec:conclusion}

This paper proposed a safety-aware Bayesian method that uses driver demonstrations to infer the preferred lateral offset and speed-dependent look-ahead distance of a pure pursuit lane keeping controller.
The look-ahead was modeled as a latent GP over speed. MCMC sampled a safety-aware inference target that combines the likelihood of closed-loop rollouts with a safety penalty on lane clearance.
The deployed controller was defined by the posterior expectations of these quantities.

The inferred controllers differed between the two drivers in both preferred path and look-ahead distance, beyond their posterior uncertainty. By reducing the posterior weight of controllers whose trajectories approached the lane boundary, the safety penalty increased the deployed controller’s clearance according to the posterior uncertainty in clearance, with little change in trajectory fit. In HIL tests with both drivers, the personalized controller ran in real time with a physical steering wheel, stayed close to its offline rollouts, and reproduced the driver’s held-out trajectories more closely than the baseline. These results show that a driver-specific, safety-aware lane keeping controller with quantified uncertainty can be inferred from demonstrations and deployed on real-time hardware without modifying the inferred parameters. However, the HIL evaluation was limited to two drivers and held-out conditions within the calibration range. Larger driver populations and driver-in-the-loop tests of driver acceptance remain for future work.

\section*{Acknowledgement}

This work was supported by the Korea Institute for Advancement of Technology (KIAT) grant funded by the Korea Government (MOTIE) (P0020536, The Competency Development Program for Industry Specialist), and by the Korea Planning \& Evaluation Institute of Industrial Technology (KEIT) grant funded by the Korean Ministry of Trade, Industry and Resources (MOTIR) (RS-2026-25544705, Development of a Scalable Heterogeneous Zonal Chassis SW Platform and Vehicle-Adaptive Drive Modules for Global OEM Customization).

\end{document}